# Near-surface ocean kinetic energy spectra and small scale intermittency from ship data in the Bay of Bengal

Jai Sukhatme, jai.goog@gmail.com, *

*Centre for Atmospheric & Oceanic Sciences and Divecha Centre for Climate Change, Indian Institute of Science, Bangalore 560012, India.*

Dipanjan Chaudhuri, dipadadachaudhuri@gmail.com

*Centre for Atmospheric & Oceanic Sciences, Indian Institute of Science, Bangalore 560012, India.*

Jennifer MacKinnon, jmackinnon@ucsd.edu

*Scripps Institution of Oceanography, University of California, San Diego, CA, USA.*

S. Shivaprasad, shivaprasad.s@incois.gov.in

*Indian National Centre for Ocean Information Services, Hyderabad 500090, India.*

Debasis Sengupta, dsen@iisc.ac.in

*Centre for Atmospheric & Oceanic Sciences and Divecha Centre for Climate Change, Indian Institute of Science, Bangalore 560012, India.*

*Corresponding author address:* Jai Sukhatme, Centre for Atmospheric & Oceanic Sciences and Divecha Centre for Climate Change, Indian Institute of Science, Bangalore 560012, India.

E-mail: jai.goog@gmail.com

# ABSTRACT

Horizontal currents in the Bay of Bengal were measured on eight cruises covering a total of 8600 *km* using a 300 *kHz* Acoustic Doppler Current Profiler (ADCP). The cruises are distributed over multiple seasons and regions of the Bay. Horizontal wavenumber spectra of these currents over depths of 12–54 *m* and wavelengths from 2–400 km were decomposed into vortical and divergent components assuming isotropy. An average of across and along track spectra over all cruises shows that the spectral slope of horizontal kinetic energy for wavelengths of 10–80 *km* scales with an exponent of $-1.7 \pm 0.05$, which transitions to steeper scaling for wavelengths above 80 *km*. The rotational component is significantly larger than the divergent component at scales greater than 80 *km*, while the two are comparable for smaller wavelengths. The measurements show a fair amount of variability and spectral levels vary between cruises by about a factor of 5 over 10–100 *km*. Velocity differences over 10–80 *km* show probability density functions and structure functions with stretched exponential behavior and anomalous scaling. Comparisons with the Garrett-Munk internal wave spectrum indicate that inertia-gravity waves account for only a small fraction of the kinetic energy except at the smallest scales. These constraints suggest that the near-surface flow in the Bay is primarily balanced and follows a forward enstrophy transfer quasigeostrophic regime for wavelengths greater than approximately 80 *km*, with a larger role for unbalanced rotating stratified turbulence and internal waves at smaller scales.

# 1. Introduction

One of the principal means of characterizing a turbulent flow is by its kinetic energy (KE) spectrum (Frisch 1995). The KE spectrum, along with other information such as the nature of small scale intermittency, the direction of interscale energy flux and the partition of KE between rotational and divergent components of the flow, constrains the possible dynamics of a fluid. Along these lines, in a quest to better understand its dynamics, the ocean's surface KE spectrum has received considerable scrutiny in both the temporal and spatial domains (Ferrari and Wunsch 2010). This has been possible through *in situ* observations, satellite altimeter data and modeling efforts.

In the spatial domain, *in situ* observations are sparse and restricted to one-dimensional (1D) cuts of the two-dimensional (2D) surface velocity field, but have the advantage of high resolution. This allows for an exploration of surface currents at scales comparable to, and significantly smaller than, the local deformation radius (Chelton et al. 1998). For example, Acoustic Doppler Current Profiler (ADCP) data from the Gulf Stream showed steep spectra close to $-3$ scaling starting near 200 *km* (Wang et al. 2010; Callies and Ferrari 2013), that flattened out to a much shallower form around 20 *km* (Callies and Ferrari 2013). Further, the rotational component of the flow was dominant down to about 20 *km* (Bühler et al. 2014), and the flattening was attributed to an increased contribution of internal wave energy at smaller scales (Callies and Ferrari 2013; Bühler et al. 2014). Similarly, cruise data from the Antarctic Circumpolar Current in the Drake Passage was observed to scale with a $-3$ power-law below 200 *km* (Rocha et al. 2016). Larger scales (200-100 *km*) were again largely rotational, while ageostrophic motions, likely from internal waves, were dominant below 40 *km* (Rocha et al. 2016). Shipboard data along 137°E in the western Pacific also indicated a transition, in terms of contribution to surface KE, from geostrophic to internal wave motions (Qiu et al. 2017). Here too, in some regions such as the Kuroshio and westward flowing North

Equatorial Current (NEC), the geostrophic or rotational modes scaled with an approximate $-3$ exponent, while the divergent component followed a shallower spectrum. The transition from geostrophic to internal waves was observed to occur at different length scales in distinct latitudinal bands, representing the diverse oceanic conditions found along this transect, (Qiu et al. 2017). Finally, drifter data from the Gulf of Mexico also suggests the dominance of divergent modes and scaling consistent with a $-5/3$ exponent at scales smaller that 5 *km*, while rotational modes had a steeper spectrum and accounted for most of the KE at larger scales (Balwada et al. 2016). Thus, the dominance of rotational modes at large scales and a significant contribution (if not dominance in some regions) of divergent modes at small scales is a consistent theme emerging from *in situ* data from different parts of the world's oceans. However, the exponents characterizing spectral slope, the ratio of rotational to divergent mode energy, and the scale at which divergent modes start contributing significantly to the total KE, all vary appreciably in ways that speak to differences in dominant physics.

Horizontal currents can also be estimated from satellite altimetry. Altimeter estimates have the advantage of longevity and global coverage, but lack the fine resolution afforded by *in situ* data. Moreover, they only allow an estimation of the geostrophic, or rotational, component of the surface ocean currents. Optimistically, scales down to approximately 100 *km* are reasonably well resolved by gridded geostrophic currents derived from the altimeter (Arbic et al. 2013; Dufau et al. 2016). Indeed, at scales between 200-100 *km*, various analyses of these geostrophic currents have reported KE spectra that follow power-laws with an approximate $-3$ exponent (Stammer 1997; Scharffenberg and Stammer 2011; Arbic et al. 2014; Khatri et al. 2018). But, it should be kept in mind that gridded currents (even above 100 *km*) suffer from noise, interpolation and smoothing, and in general, gridded spectra tend to be steeper than those from raw satellite track data (Arbic et al. 2013; Khatri et al. 2018; Wortham and Wunsch 2014; Xu and Fu 2012). As with *in situ* data, spectral exponents estimated from altimetry show regional dependency (Xu and Fu 2012; Zhou et al. 2015). In some cases flatter spectra have been reported (LeTraon et al. 2008), which may reflect a contribution from ageostrophic motions (Zhou et al. 2015; Richman et al. 2012). Given the caveats regarding KE spectra from altimeter currents, the most exciting aspect of this data is the ability to estimate spectral energy fluxes (Scott and Wang 2005). Specifically, scales larger than approximately 200 $km$ show an inverse transfer of KE (Scott and Wang 2005; Tulloch et al. 2011; Arbic et al. 2014; Khatri et al. 2018), while those between 200 and 100 $km$ exhibit a robust forward transfer of enstrophy which is consistent with a $-3$ KE spectrum (Khatri et al. 2018). Fluxes have also been estimated in the frequency domain, and show that the inverse transfer of KE at large scales is accompanied by a simultaneous transfer to low frequencies (Arbic et al. 2014).

In this context, high resolution eddy resolving numerical ocean simulations, both regional and global, have helped clarify the ageostrophic nature of forward energy transfers at submesoscales (Capet et al. 2008), confirmed the flattening of spectra due to internal waves near tide generating regions (Richman et al. 2012), yielded exponents close to $-3$ for geostrophic currents (Khatri et al. 2018; Biri et al. 2016) and shown a steep to shallow transition in scaling associated with rotational and divergent components of the surface flow (Rocha et al. 2016). Further, inverse KE and forward enstrophy transfer regimes have also been observed in models, and smoothening exercises have verified their qualitative robustness (Khatri et al. 2018; Arbic et al. 2014).

Here, we contribute to the study of spectra from *in situ* observations by analyzing ADCP currents from the Bay of Bengal. The upper ocean in the Bay of Bengal is distinguished by unusually shallow mixed layers and large lateral gradients, which can give rise to an energetic submesoscale (Lucas et al. 2016; Ramachandran et al. 2018). Improved understanding of upper ocean dynmics is also of particular interest in the Bay of Bengal, due to the important role the upper ocean plays in mediating the summer monsoon via air-sea interactions (Mahadevan et al. 2016). As far as we are aware, high spatial resolution statistical characteristics of surface currents from this region have not been reported before. We examine ship-borne ADCP observations collected on several research cruises in the Bay of Bengal during the period 2014-2017, with each of the cruise tracks covering a distance of at least 900 *km* in the open ocean. The geographical setting and details of the data are presented in Section II. Spatial KE spectra, the contribution of rotational and divergent components, Garrett-Munk spectra, structure functions and probability density functions of across and along track velocity increments are presented in Section III. We then interpret these observational results in terms of the implied constraints on possible dynamical regimes at work in the near-surface layer of the Bay of Bengal. Finally, a discussion in the broader context of rotating and stratified turbulence and the upper tropospheric Nastrom-Gage spectrum concludes the manuscript.

## 2. Oceanographic Setting and Data

The ship tracks used in this study are shown in Figure 1. The tracks range from approximately 8°N to 20°N in latitude and 81°E to 90°E in longitude. Details of the cruises, including their dates, track length and error estimates are provided in Table 1. The upper ocean velocity was measured by a 300 *kHz* ADCP mounted at on the hull of Research Vessel Sagar Nidhi (SN) at a depth of 5 *m* below the sea surface. The sampling rate was set to 1 *Hz*, and sampling depths were set to 2 *m* vertical bins. The ADCP data was processed using the RDI software (http://rdinstruments.com). We obtained absolute ocean velocity by removing the ship's velocity, calculated using navigation inputs from a differential Global Positioning System (GPS), from the total velocities. The ADCP was configured to record 1 *min* averages, which corresponds to a horizontal spatial resolution of 120-250 *m*, depending on ship speed. The velocity data was then interpolated onto a uniform 1 *km* horizontal grid using cubic splines. Finally, the ADCP velocities were rotated to along-track (a frame of reference aligned with the ship track) and cross-track coordinates, for analysis. There are total of eight tracks, each of which measures currents at twenty four levels ranging from 12 to 54 *m*. Thus, in all we have 192 segments of a few hundred *km* in length. More details on data collection, processing and error estimation are presented in the Appendix A1–A4. In some cases, the ship followed a convoluted track, and we do not consider these looped portions in our analysis.

## 3. Results

Currents from a representative section (SN120) are shown in Figure 2. This cruise passed through two eddies, visible in the geostrophic currents shown in Figure 2a. The zonal and meridional currents reflect this via a long approximately 600 *km* oscillation visible in Figures 2b and 2c, which by eye is vertically coherent over the measurement range. The first baroclinic deformation scale in the Bay ranges from about 60 to 150 *km* (Chelton et al. 1998). Eddy length scales estimated from altimeter data also lie in this range, though towards the larger end of these estimates (Paul and Sukhatme 2019). Using a typical length scale of 100 *km*, and the RMS velocity along each cruise, the Rossby number associated with this mesoscale eddy is approximately 0.07. Currents at any one depth reveal considerable small-scale variability superimposed on the broad mesoscale structure (Figure 2d). Our intent in the following Section is to statistically characterize this small scale variability in the near-surface currents.

Density stratification is estimated using uCTD data where available (SN88 and SN100), and in the other cases, from monthly temperature and salinity data from the Roemmich-Gilson 1° gridded

Argo dataset (Roemmich and Gilson 2009). Specifically, we take a 2° wide strip centered on each ship-track, and plot mean salinity and temperature profiles from the Argo dataset (see Appendix A6; Figure A4). This data suggests that (i) cruise-mean upper mixed layer depth is between 10 and 25 $m$, and (ii) density in the upper upper 50 $m$ is mainly determined by salinity (Sengupta et al. 2016), with some contribution from temperature, particularly in March 2017 (Cruise SN115). Mean profiles, averaged over all cruises, of salinity and temperature are shown in Figure 3. In the depth interval 12-54 $m$ that we have considered for computing velocity spectra, the average stratification (Brunt-Vaisala frequency squared) over all cruises is approximately $2 \times 10^{-4} s^{-2}$ (see Appendix A6; Figure A4 for individual cruises) and the mean Froude number is 0.4. Some caution needs to be exercised when interpreting the Froude number as the stratification varies within the depth where the currents are measured. Thus, based on the Rossby and Froude numbers, the surface currents in the Bay are in a regime of rapid rotation and moderately strong stratification.

*a. Spectra*

Figure 4 shows the mean horizontal KE (HKE) spectrum obtained by averaging spectra from individual cruises. The spectra for individual cruises are shown in Figures A5 and A6, and details on computation of spectra and error estimates can be found in Appendix A1–A4. As mentioned, the data consists of currents at twenty four depths from 12 to 54 $m$. Spectra are computed for every depth and their mean represents an average spectrum for a given cruise. Note that spectra at different depths do not vary significantly (Appendix A5), as is exemplified by cruise SN115 shown in Figure A3. Hence, in this work, we consider mean spectra as averaged over all levels.

Several patterns emerge from the average across and along track spectra Figure 4a. In particular, we observe a shallowing of spectral slope with higher wavenumbers. The slope as seen by eye is close to $-5/3$, and in fact, a best fit over 10–80 *km* yields an exponent of $-1.7 \pm 0.05$ (best fits for individual cruises are presented in Table A1). Further, even though there are only a few data points at large scales, the slope has a tendency to become steeper for scales greater than 80 *km*. Indeed, a best fit over 80–300 *km* yields an exponent of $-2.15 \pm 0.17$. A good example of a change in slopes is SN120 (Figure A5d), where the HKE scales with an exponent of almost $-3$ ($-1.7$) above (below) 80 *km*. Overall, this change in scaling and the power-law exponents are better seen in compensated spectra, i.e., spectrum multiplied by $k^2$, shown in Figure 4b. Clearly the along and across track spectra are 'blue' at scales smaller than (wavenumbers larger than) approximately 80 *km*. In most cases, the blue-er spectral slopes extend to the highest wavenumbers resolved here, but we restrict our attention to scales greater than 10 *km*.

Additional insight comes from separation of rotational (geostrophically balanced) and divergent (unbalanced) motions. The two types of spectra can be extracted from along and cross-track spectra following the related methods of Bühler et al. 2014 and Lindborg 2015. The results from both methods are shown in Figure 4c. Both approaches assume isotropy and homogeneity, but the mathematical techniques involved and their numerical implementation lead to differing estimates of rotational and divergent spectra. A comparison and merits of each approach are discussed by Bierdel et al. (2016), and it should be kep tin mind that these methods come with significant caveats in their ability to interpret two-dimensional flow fields from one-dimensional slices (Shcherbina et al. 2013).

As seen in Figure 4c, from 10–80 *km*, the slope of the rotational and divergent spectra is comparable. The closeness of these estimates, in this scale range, are markedly different from results reported by Bühler et al. 2014 . In addition, quite noticeably, the divergent spectrum remains shallow even at large scales (i.e., greater than 80 *km*). Further, the change in slope of the across and along track spectra is accompanied by a gradual change from steep to shallow scaling in the rotational spectrum. Thus, it is possible that the shallowing of KE spectra at higher wavenumbers is due to both an increasing contribution of divergent modes, as well as a transition in the scaling of the rotational modes themselves. From Figure 4d, we note that the rotational energy is greater then the divergent contribution at scales larger than approximately 80 *km*, while the two are roughly equal over the range of 10–80 *km*. Thus, the range of 10–80 *km* is one where the spectral slope is relatively shallow and both the rotational and divergent modes contribute to the surface KE.

Qualitatively, the greater energy in rotational modes at large scales, and transition to a regime where the divergent modes play an important role, is in line with *in situ* observations from the Gulf Stream, Drake Passage and the western Pacific Ocean (Bühler et al. 2014; Rocha et al. 2016; Qiu et al. 2017). Moreover, divergent modes become significant in an energetic sense at approximately 80 *km*, which is consistent with the expectations from high resolution simulation estimates of the transition scale in this region (Qiu et al. 2018). Together with the fact that our measurements are well resolved down to 10 *km*, we observe shallow scaling for the scales where both rotational and divergent modes contribute to the KE spectrum. Also, the tendency towards steeper scaling at larger scales (above 80 *km*) ties in well with altimeter estimates of the spectrum in the range 200-100 *km* from the Bay of Bengal (Paul and Sukhatme 2019).

Some of the divergent component may be due to internal waves, which enter these estimates through a combination of their horizontal wavelengths and frequency time aliasing. In general, we expect to see a spectral signal from both energetic low-mode internal tides known to propagate through the Bay (e.g. Wijesekera et al. 2016, Lucas et al. 2016), and an internal wave continuum. The latter is often represented with a canonical Garrett-Munk (GM) spectrum (Garrett and Munk 1972; Gregg and Kunze 1991). Some previous studies have found observed horizontal KE spectra to be well described by the GM internal wave spectrum, particularly at higher wavenumbers (Klymak et al. 2015). To assess the role of internal waves, following Klymak et al 2015, we have added GM spectra to Figure 4a. Here, we use mean stratification values for all cruises via a combination of uCTD (SN88 and SN100) and Argo data, averaged over the depth range of the currents (GM estimates for invidual cruises are shown in Figures A5 and A6).

Between 10 and 80 *km*, the GM continuum levels are well below the measurements, and other processes must be at play. Given their abundance, it is likely that near-inertial oscillations (Johnston et al. 2016) and internal tides (Murty and Henry 1983; Mohanty et al. 2018), neither of which is well represented by the GM spectrum, may account for a fair portion of the divergent component in the Bay. Indeed, there are locations in the Bay where the tidal displacement signal is larger than the inertia-gravity GM spectrum (Wijesekera et al. 2016). Note that, in a few cruises (Figure A5, especially SN88 and SN100), the GM line agrees well with the measured divergent component of the spectrum, but only at wavelengths smaller than 10 *km*.

*b. Small scale intermittency*

The spectrum only reflects the second moment of fluctuations, and higher moments are required for a more quantitative estimate of intermittency in the underlying field. The hope is that these statistics will help to further constrain the surface dynamics of the Bay. In this subsection, we will focus on scales between 10-80 *km*. In particular, we now turn to structure functions of velocity increments. For a one-dimensional field, these are defined as (Frisch 1995),

$$S_q(r) = \langle |u(x+r) - u(x)|^q \rangle, \tag{1}$$

where, *u* is the across or along track velocity, *r* is a spatial increment, *q* is the order of the structure function and $\langle \cdot \rangle$ denotes an average over all possible realizations for a given *r*. If $S_q(r)$ scales as a power-law with $r$ (over a specified range) then the scaling exponents ($\zeta_q$) are defined as (Frisch 1995; Sreenivasan and Antonia 1997),

$$S_q(r) \sim r^{\zeta_q}, r_1 \leq r \leq r_2. \qquad (2)$$

Here, $r_1, r_2$ are the inner and outer scales over which the power-law behavior of $S_q(r)$ holds true. By definition, the exponent ($n$) of the power spectrum is given by, $n = -(1+\zeta_2)$, for $n > -3$. Larger $q$ highlight extremes in the signal, and a nonlinear dependence of $\zeta_q$ on $q$, i.e., $\zeta_q \neq q\zeta_1$, is referred to as anomalous scaling (Frisch 1995; Sreenivasan and Antonia 1997). Scaling exponents of across and along track data, for increments in the range 10 to 80 *km* — i.e., $r_1 = 10$ and $r_2 = 80$ *km* in (2), are shown in Figure 5. The plots are an average of $\zeta_q$ obtained as slopes of best fit lines to log-log plots of (2) from different ship tracks. There are slight differences in the scaling exponents in the along and across track directions, especially for higher moments, but these lie within the error bars of estimates from different tracks. Note that $\zeta_2 \approx 2/3$, which is consistent with the $-1.7$ power spectrum exponent over this range of scales. Also, $\zeta_q$ is a nonlinear function of $q$, and rises much slower than $q\zeta_1$. Thus, increments of ADCP surface currents exhibit anomalous scaling and are expected to be highly intermittent between scales of 10 and 80 *km*. This anomalous scaling is likely to manifest itself in the multifractal nature of fields based on derivatives of the velocity field (Paladin and Vulipiani 1987).

As seen in Figure 5, the error bars on scaling exponents become larger with $q$. Indeed, as a few extreme samples dominate (1), it becomes increasingly difficult to get an accurate estimate of $\zeta_q$ for large $q$. Rather than individual moments, it is often advantageous to compute the probability density function (PDF) of the field (She et al. 1988; Kailasnath et al. 1992). As with the structure functions, the shape of the PDF yields insight into the statistical character of the process. The PDF of across and along track velocity increments ($10 \leq r \leq 80$ *km*, normalized by their root mean square value) is shown in the first panel of Figure 6, and is non-Gaussian in nature. In fact, as is the case in the forward energy transfer regime of three-dimensional homogeneous turbulence (She et al. 1988; Kailasnath et al. 1992; Noullez et al. 1997), the PDF is better described as stretched exponential with a much larger probability of encountering extremes than for a normal distribution. Recent work employing drifter data from the Gulf of Mexico shows similar non-normal distributions of velocity increments over short distances (Poje et al. 2017). Further, PDFs for increments in two different ranges $10 \leq r < 30$ *km* and $50 \leq r < 70$ *km* are shown separately in the second panel of Figure 6. Here, an increase in degree of intermittency, or the more stretched nature of the distributions with a narrower and taller peak is noticeable for smaller increments (Noullez et al. 1997). In fact, this non-self similarity of distributions for different increments, or the scale dependent nature of intermittency, is tied to the anomalous scaling of structure functions (Biferale 2003).

## 4. Interpretation

We are now in a position to ask, does this information on spectra and intermittency help constrain the possible dynamical regimes at work on the surface of the Bay? Above 80 *km*, the small Rossby number (approximately 0.07 at 100 *km*), much larger energy in rotational modes and signs of a steep spectrum on average (and in particular for some cruises as seen in Figures A5 and A6) suggest a quasigeostrophic (QG) forward enstrophy transfer regime (Charney 1971), much like what was explicitly demonstrated by calculating fluxes using altimeter data in the range of 200-100 *km* in other oceanic regions (Khatri et al. 2018).

At smaller scales, i.e.,10 *km* to 80 *km*, the spectral slopes become shallower. The (relatively) increasing energy level of higher wavenumber motions has several possible dynamical explanations.

Surface quasigeostrophic (SQG) theory (Blumen 1978; Held et al. 1995; Lapeyre 2017) is sometimes invoked to understand low Rossby number rotating and stratified Boussinesq flows near the presence of a boundary or rapid change in stratification. In our observations, from 10-80 *km*, the Rossby number is still quite small (approximately 0.3 at about 25 *km*), and SQG theory may be appropriate. The $-5/3$ spectral slope predicted by this theory, particularly in the upper ocean, is within error bars of the observed mean spectral slope as well as that of several individual cruises. Keeping in mind that surface temperature/buoyancy scales in a similar manner as the velocity field for SQG dynamics, the anomalous scaling of temperature structure functions in SQG (Sukhatme and Pierrehumbert 2002), is in accord with the intermittency of ADCP currents at smaller scales.

Taken together, this view suggests the flow on the surface of the Bay is balanced and follows a QG forward enstrophy transfer regime at large scales (above 80 *km*) (Charney 1971), and SQG dynamics at smaller scales ($-5/3$ scaling, 80–10 *km*) (Pierrehumbert et al. 1995). Indeed, the expected shallow scaling of surface KE in SQG dynamics (Pierrehumbert et al. 1995), has been observed in numerous simulations (Okhitani and Yamada 1997; Sukhatme and Pierrehumbert 2002; Celani et al. 2004; Capet et al. 2008). But, if balanced SQG dominates the dynamics at small scales, then we expect most of the energy to be in the rotational component of the flow, which is at odds with observation that the divergent part is of comparable magnitude in the 10-80 *km* wavelength range.

Internal wave contributions may play an important role here. Additionally, the presense of other types of ageostrophic motions has been noted in high resolution studies aimed at the upper ocean (Klein et al. 2008), and more generally in idealized studies of unbalanced rotating-stratified turbulence (Bartello 1995; Sukhatme and Smith 2008; Vallgren et al. 2011; Deusebio et al. 2013; Kafiabad and Bartello 2016; Asselin et al. 2018). At small scales, along with a $-5/3$ exponent, this regime provides a route for dissipation via a robust forward transfer of KE (Molemaker et al. 2010; Deusebio et al. 2013; Nadiga 2014; McWilliams 2016).

It is important to note that there is a significant change in the behavior of predicted or modeled unbalanced rotating-stratified turbulence with Rossby number. In particular, at low Rossby numbers (i.e., $<0.1$), there is a scale below which divergent modes account for a larger fraction of the KE than the rotational modes, and this transition from rotational to divergent mode dominance is what leads to a change in the exponent of the KE spectrum ($-3$ to $-5/3$ scaling) (Nadiga 2014; Kafiabad and Bartello 2016; Asselin et al. 2018)[1].

On the other hand, for moderate rotation rates (i.e., Rossby number $\gtrapprox 0.1$), along with the $-5/3$ scaling of divergent modes, the rotational spectrum also shows shallow scaling, and similar to purely stratified 3D turbulence (Lindborg and Brethouwer 2007), there is a roughly equal contribution of the divergent and rotational modes to the KE (Deusebio et al. 2013). Thus, for Rossby numbers $\gtrapprox 0.1$, much like what we observe at 10-80 *km* in the Bay, a shallow spectrum emerges due to both an increasing contribution of divergent modes as well as a shallowing of the rotational spectrum itself (Deusebio et al. 2013).

As a whole, this suggests that at scales greater than approximately 80 *km* (with Rossby number $\approx 0.07$), the surface flow in the Bay is balanced and follows a QG forward enstrophy transfer regime, while at smaller scales (down to 10 *km*), the regime is one of unbalanced rotating-stratified turbulence (with Rossby number $\gtrapprox 0.1$) that merges smoothly into stratified turbulence at progressively smaller scales.

---

[1] Note that at very small Rossby numbers, i.e. $< 0.01$, unbalanced modes have very little energy, and there is almost no transfer of KE to small scales (Kafiabad and Bartello 2016).

## 5. Discussion and Concluions

From a broader perspective, the study of rotating and stratified turbulence, as well as rotational and divergent, or balanced and unbalanced portions of geophysical flows, has a significant atmospheric literature that may provide some insight here. In the atmosphere, analogous studies have been been motivated by a quest to understand the Nastrom-Gage spectrum (Nastrom and Gage 1985). This spectrum, estimated from aircraft data, represents the upper tropospheric KE and is observed to follow a power-law that transitions from a $-3$ to $-5/3$ exponent. The change in scaling takes place at approximately 500 *km*, with steeper slope at large scales and shallower slope at smaller scales (Nastrom and Gage 1985). The large scale $-3$ spectrum is quite widely accepted to be the rotationally dominated enstrophy cascading regime of QG turbulence (Charney 1971). The smaller or mesoscale $-5/3$ scaling has proved to be more tricky. In idealized settings, some efforts emphasize the emergence and energetic dominance of unbalanced or divergent mode turbulence (for example, Kitamura and Matsuda 2006; Vallgren et al. 2011; Nadiga 2014; Asselin et al. 2018), while others have pursued a purely balanced framework for the mesoscale regime (Tulloch and Smith 2006).

Given the fact that it can help to decide between the competing theories, a recurring theme is the relative contribution of rotational and divergent modes to the mesoscale KE (Cho et al. 1999; Lindborg 2007; Callies et al. 2014; Lindborg 2015). In fact, recent analysis of aircraft data suggests that the ratio of rotational to divergent mode energy is not universal, but is dependent on latitude and height (Bierdel et al. 2015). Further, mesoscale range KE might be influenced by factors outside idealized scenarios; for example, contributions of moist convection to the $-5/3$ scaling (Sun et al. 2017). Indeed, the fact that spatially uncorrelated initial data can spontaneously given rise to such scaling has been noted in a moist setting (Sukhatme et al. 2012).

Though many facets of the dynamics are quite different, the similarity between the flow near the tropopause and surface ocean current measurements reported here is intriguing. In both cases, KE spectra are steep at large scales and then flatten to a shallower scaling at smaller scales. Here, large and small scales are with respect to the local deformation radius. Further, these relatively large scales are balanced, dominated by rotational modes, and the scaling of KE follows the enstrophy cascading regime of QG turbulence. Indeed, this is seen in altimeter fluxes, eddy resolving models as well as via spectra of *in situ* ADCP based data. In both situations, divergent modes play an increasingly large role at smaller scales.

The transition scale between these two regimes, and the actual ratio of rotational to divergent energy appears to be dependent on geography and local environmental setting. Our analysis shows that in the Bay of Bengal this transition occurs at approximately 80 *km*, i.e., on average, rotational modes have more energy above 80 *km*, and from 80-10 *km*, divergent modes contribute significantly and have about the same KE as the rotational component. Moreover, from 80-10 *km*, the observed anomalous scaling of velocity increments observed in the ocean data presented here is consistent with three-dimensional stratified turbulence in other geophysical fluids (Lohse and Xia 2010). Specifically, *in situ* measurements of stratified turbulence, for example, through marine clouds (Seibert et al. 2010) and the atmospheric surface layer (Chu et al. 1996) also show anomalous scaling and non-Gaussian distributions of velocity increments.

The results presented here offer a first look at energy levels and suggestive dynamical balances at different length scales in the near-surface flow of the Bay of Bengal. The overall impression is one of QG-like balanced motion at large scales, and increasingly divergent motions below 80 *km* wavelengths, qualitatively similar in some ways to atmospheric near tropopause measurements. The divergent spectra reported here may reflect a combination of the forward KE cascade of strongly ageostrophic motions and a variety of internal waves. Further efforts can explore the role of strong seasonality, the distinct geography of postmonsoonal fresh water inflow and associated enhanced submesoscale activity (Sengupta et al. 2016; Sarkar et al. 2016; Ramachandran et al. 2018).

*Acknowledgments.* The authors are grateful to the Ministry of Earth Sciences, New Delhi for support through the National Monsoon Mission, Indian Institute of Tropical Meteorology (IITM), Pune. JS would also like to acknowledge support from the Department of Science and Technology, under project number DSTO1633 and the University Grants Commision (UGC) under the Indo-Israel Joint Research Program ($4^{th}$ cycle) for project F 6-3/2018. JAM would also like to gratefully acknowledge support from the US Office of Naval Research, grant number N00014-14-1-0455.

# APPENDIX

## Collecting and Processing of ADCP data

### A1. Error budget

A 300 kHz ADCP mounted on the hull of the RV Sagar Nidhi measured velocity profiles in the upper $12-80m$ of the ocean, using navigation inputs from a differential GPS system and processed by "WinADCP software" (http://www.teledynemarine.com/rdi/). On all cruises, we have gap-free data in the depth range $12-54m$. An estimate of "error velocity" is obtained from the redundant information among the four beams of the ADCP The estimated accuracy of the measured relative velocity by the RDI ADCP is $\sigma_{rel} \sim 10 \times 10^{-3} m/s$. The accuracy of the GPS system is $\sigma_{vel} \sim 2.5 \times 10^{-3} m/s$. Hence, the total estimated random error of the 1 *min* ADCP velocity is,

$$\sigma_{ADCP} = \sqrt{\sigma_{rel}^2 + \sigma_{vel}^2} \approx 10.11 \times 10^{-3} m/s. \tag{A1}$$

## A2. Uniform Grid

The horizontal resolution of raw ADCP velocity is 120-250 *m*, depending on ship speed. To make uniform 1 *km* horizontal grid we follow the following steps: (i) First we smooth 1 *min* horizontal velocities by a $5-point$ moving average. (ii) Then we interpolate (hereafter is known as INTPOL) the smooth velocity onto a uniform 1 *km* horizontal grid by cubic spline. We also make gridded data (hereafter AVERG) by taking the mean of all the data points ($4-5$ points on average) that lie within 1 *km* grid along the path of the ship. Figure A1 shows that the KE (from cruise SN112) of the gridded currents do not differ significantly at scales larger than 5 *km*, when we change the averaging methods from INTPOL to AVERG. Other crusises considered in this work show a similar behavior.

## A3. Signal-to-noise ratio

The ratio of signal power to the noise power is known as signal-to-noise ratio (SNR). It is an indicator of the robustness of the signal compared to the noise power, and reads,

$$SNR = \frac{V_{signal}^2}{V_{noise}^2}, \qquad (A2)$$

where *V* represents the root-mean-square amplitude. For the cruise SN100, the noise floor starts at 3.1 *km* (Figure A2); $SNR_{SN100} = (V_{10-800km}/V_{2-3.1km})^2$ becomes $2 \times 10^4$. The SNR values for the 1 *km* gridded data from eight cruises estimated using equation A2, lie between $0.3 \times 10^4 - 30 \times 10^4$, these high SNR values permit us to carry out analysis with 1 *km* gridded data.

## A4. Confidence limits in spectral estimates

Currents from twenty-four levels ranging from 12-54 *m* depth are used to calculate spectra in this study. First, the spectrum of each level is obtained using "pwelch" command in Matlab with 400 *km* Hamming window and 50% overlap. Following Percival and Walden, 1993, a good estimate of the degrees of freedom (DOF) of spectra estimated using 50% overlapped segments in Welch's overlapped segment averaging (WOSA) approach is $\nu = \frac{36 \times N_B^2}{19N_B - 1}$. $N_B$ is the total number of blocks to be averaged. In our case $N_B$ is 2. The velocities are highly vertically correlated for ADCP measurements from each of the cruises. We consider only one independent realisation of the spectra from each of the cruises with 4 DOF. The mean spectrum is obtained by averaging spectra from eight legs from different cruises. Hence, the final DOF becomes $\nu \times 8 = 32$. The error bar is estimated by considering this DOF and a 95% confidence interval of a $\chi$-square distribution.

## A5. Spectra at different depths (SN115)

Figure A3 shows the HKE spectra at different depths for cruise SN115. In both the along and across track directions, quite clearly, the spectra do not vary significantly with depth. This lack of depth dependence is also seen in other cruises. Hence, in this work, we deal with depth averaged HKE spectra.

## A6. Spectra and T,S and buoyancy profiles from individual cruises

The temperature, salinity and buoyancy profiles for each cruise are shown in Figure A4. Here, density stratification is estimated from uCTD (SN88 and SN100) and Roemmich-Gilson 1° gridded Argo data. When averaged over 12 to 54 *m*, the buoyancy frequency varies from 1–5 $\times 10^{-4} s^{-2}$, with a mean of $2 \times 10^{-4} s^{-2}$.

Depth averaged across, along track spectra with GM estimates for each cruise is shown in Figures A5 and A6. As with the average picture described in the main text, most cruises show power-law that has shallow (steep) scaling at small (large) scales. Clearly, the transition scale varies from cruise to cruise, and works out to approximately 80 *km* when averaged over all tracks. In the smaller range of scales, the best fit estimates of the slopes, and the range over which they are estimated are presented in Table A1.

# LIST OF TABLES

TABLE 1. Cruise numbers, duration, track length, and error estimates. Cruises SN110 and SN112 have been subdivided into two legs for analysis.

| Cruise | Start Date | End Date | Track Length (*km*) | Error (*mm/s*) |
|---|---|---|---|---|
| SN88 | 26 August 2014 | 31 August 2014 | 890 | ±1 |
| SN100 | 24 August 2015 | 01 September 2015 | 1500 | ±2 |
| SN110A | 19 December 2016 | 22 December 2016 | 1000 | ±2 |
| SN110B | 28 December 2016 | 02 January 2017 | 900 | ±4 |
| SN112A | 03 February 2017 | 07 February 2017 | 1000 | ±6 |
| SN112B | 09 February 2017 | 17 February 2017 | 1100 | ±10 |
| SN115 | 19 March 2017 | 22 March 2017 | 1000 | ±38 |
| SN120 | 02 June 2017 | 08 June 2017 | 1300 | ±14 |

Table A1. Spectra slopes and range of scales found on each of the cruises. The best-fit spectral slopes are indicated in parentheses.

| Cruise | Across-Track | Along-Track |
|---|---|---|
| SN88 | $-2$ ($-1.9 \pm 0.12$); 10-80 km | $-2$ ($-2.17 \pm 0.15$); 10-80 km |
| SN100 | $-5/3$ ($-1.58 \pm 0.15$); 10-80 km | $-5/3$ ($-1.98 \pm 0.07$); 10-80 km |
| SN110A | $-2$ ($-2.47 \pm 0.10$); 10-80 km | $-2$ ($-2.34 \pm 0.1$); 40-80 km |
| SN110B | $-2$ ($-2.42 \pm 0.1$); 10-80 km | $-2$ ($-2.03 \pm 0.1$); 10-80 km |
| SN112A | $-5/3$ ($-1.88 \pm 0.1$); 10-80 km | $-5/3$ ($-1.85 \pm 0.1$); 10-80 km |
| SN112B | $-5/3$ ($-1.72 \pm 0.4$); 40-80 km | $-5/3$ ($-1.77 \pm 0.4$); 30-80 km |
| SN115 | $-5/3$ ($-1.71 \pm 0.10$); 10-80 km | $-5/3$ ($-1.72 \pm 0.11$); 30-80 km |
| SN120 | $-5/3$ ($-1.71 \pm 0.5$); 30-80 km | $-5/3$ ($-1.23 \pm 0.5$); 30-80 km |

# LIST OF FIGURES

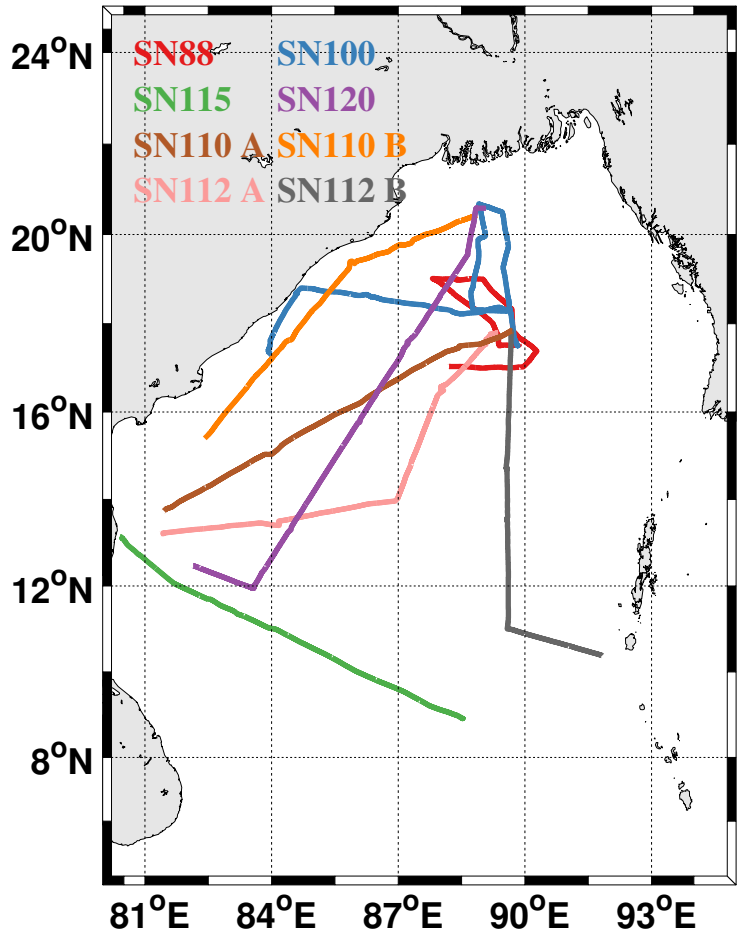

FIG. 1. Ship tracks in the Bay of Bengal used in this study. Details of the dates, track lengths and error estimates can be found in Table 1.

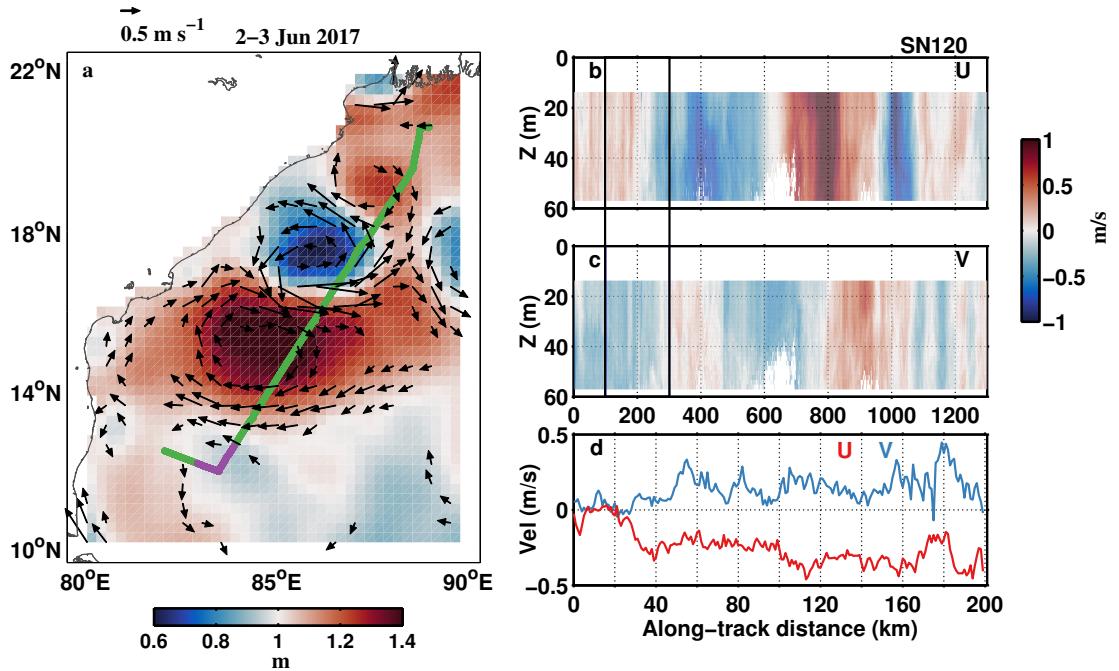

FIG. 2. Panel (a) SN120 track, geostrophic currents and height (from AVISO). Panel (b) Zonal and (c) meridional currents, at all depths from 12 to 54 *m* over approximately 1200 *km* at a resolution of 1 *km*. Panel (d) Zoom into the currents over a 200 *km* stretch marked in panels (b,c). Though the entire leg of the cruise is shown, calculations of spectra are restricted to the portion where the ship followed a straight line.

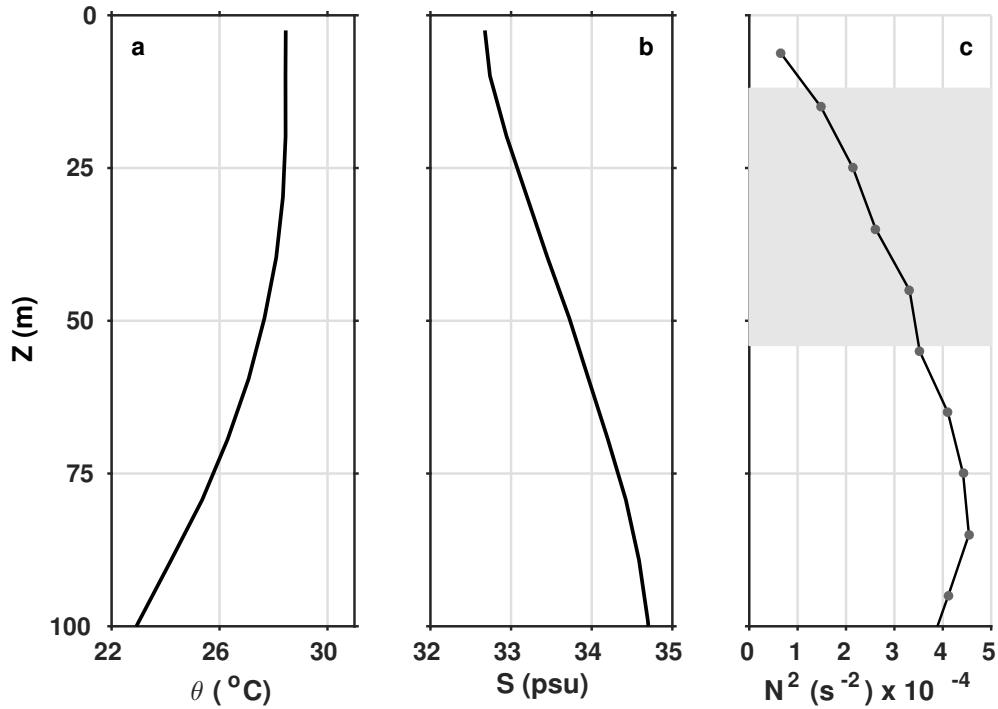

FIG. 3. Mean profiles of (a) temperature ($\theta$), (b) salinity ($S$), and (c) Brunt-Vaisala frequency squared ($N^2$) from gridded Argo dataset near the cruise tracks. Light gray background in panel (c) stands for 300 kHz ADCP's measurement depth, i.e. $12-54$ m.

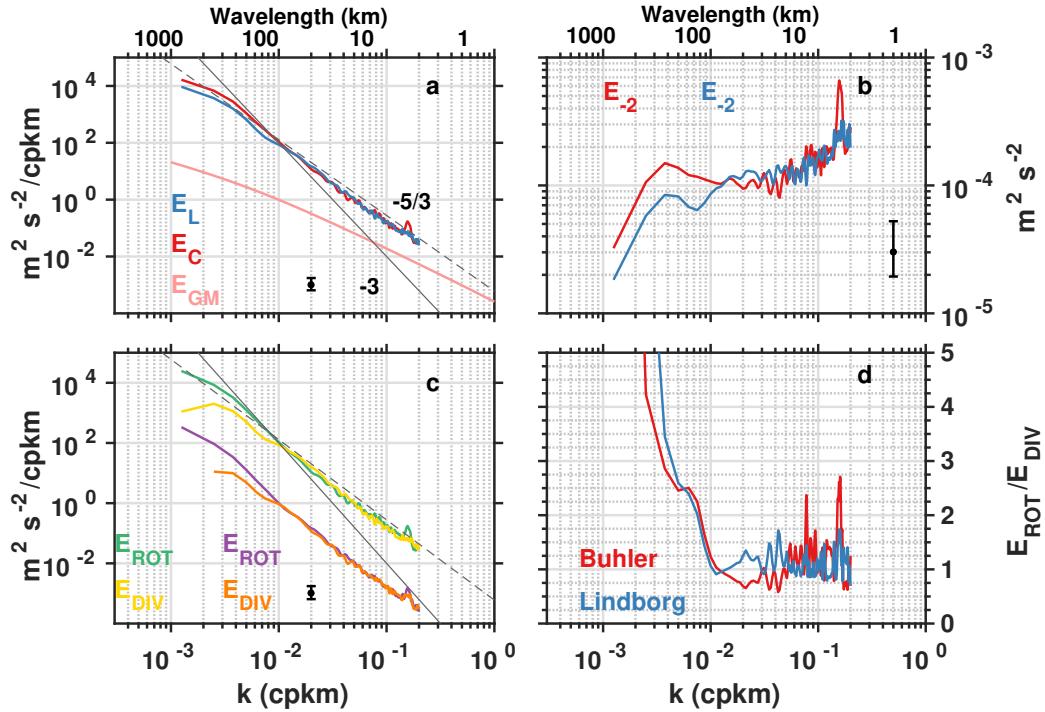

FIG. 4. (a) Mean along-track ($E_L$; blue), cross-track ($E_C$; red) HKE spectra, and an estimate of the Garret-Munk ($E_GM$; pink) spectrum. (b) Mean compensated ($E_{-2}$) along-track (blue) and cross-track (red) spectra. (c) Rotational ($E_{ROT}$) and divergent ($E_{DIV}$) spectra estimated by the methods of Bühler et al. 2014 (green and yellow) and Lindborg 2015 (purple and orange). Rotational and divergent spectra estimated using Bühler and Lindborg are offset by 0.1. (d) The ratio of rotational to divergent energy estimated by Bühler (red) and Lindborg (blue) methods. The slanted gray line represents the slope of $-3$ and the slanted gray dashed line represents a slope of $-5/3$ in the panels (a) and (c). Error bars represent the confidence limits in (a)-(c).

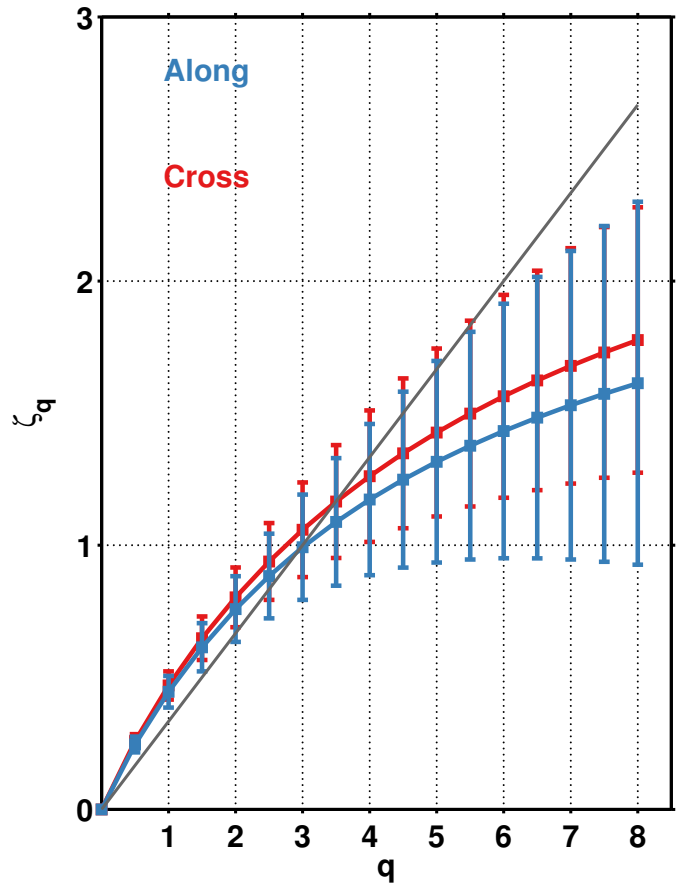

FIG. 5. Scaling exponents $\zeta_q$ vs the moment order $q$. These represent a mean over all shiptracks. Error bars are derived by estimates from individual cruises. The straight line for reference is $q/3$.

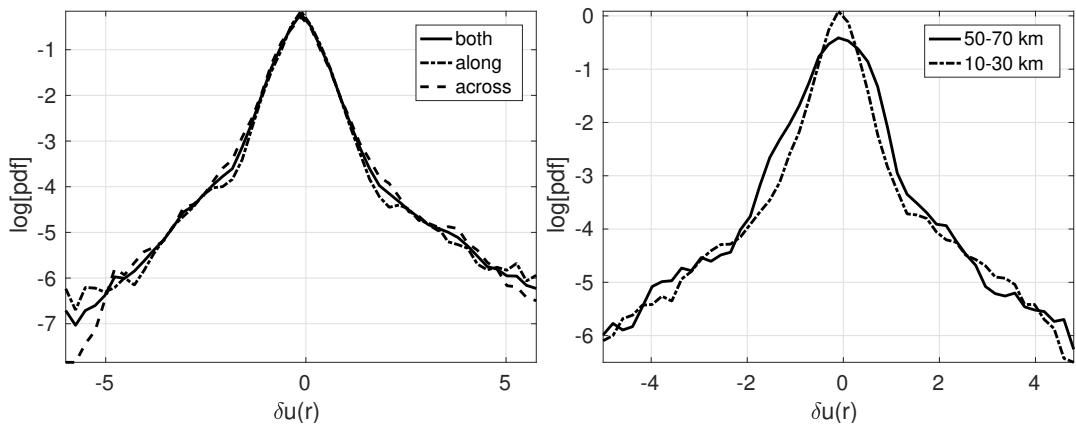

FIG. 6. Probability density functions of $\delta u(r) = u(x+r) - u(x)/\langle [u(x+r) - u(x)]^2 \rangle^{1/2}$, i.e., the velocity increments normalized by their root mean square value. (a) Across (dash-dot), along track (dash) and both together (solid). (b) Across and along track together for increments in the range 10-30 *km* (dash-dot) and 50-70 *km* (solid). The distributions follow a stretched exponential form, i.e., $\exp[-\alpha(\delta u)^\beta]$; with both parameters $\alpha, \beta$ being functions of *r*.

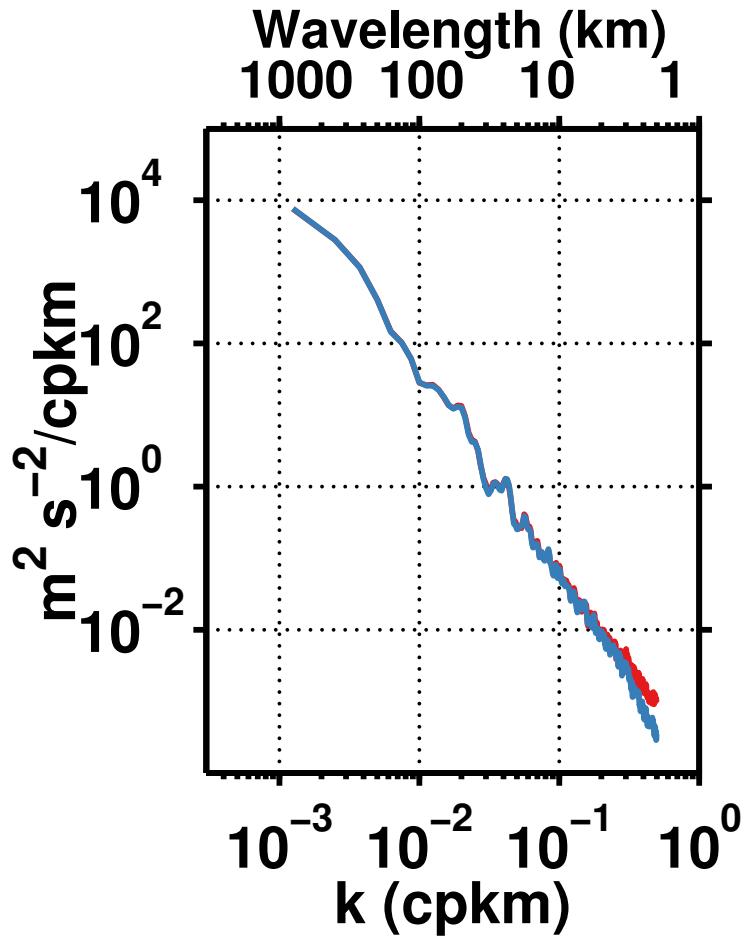

Fig. A1. Spectra of 12 − 54 m depth-averaged velocity magnitude from 300 kHz ADCP for SN112, calculated over 1 km grid by cubic-interpolation (blue) and block-averaging (red) method.

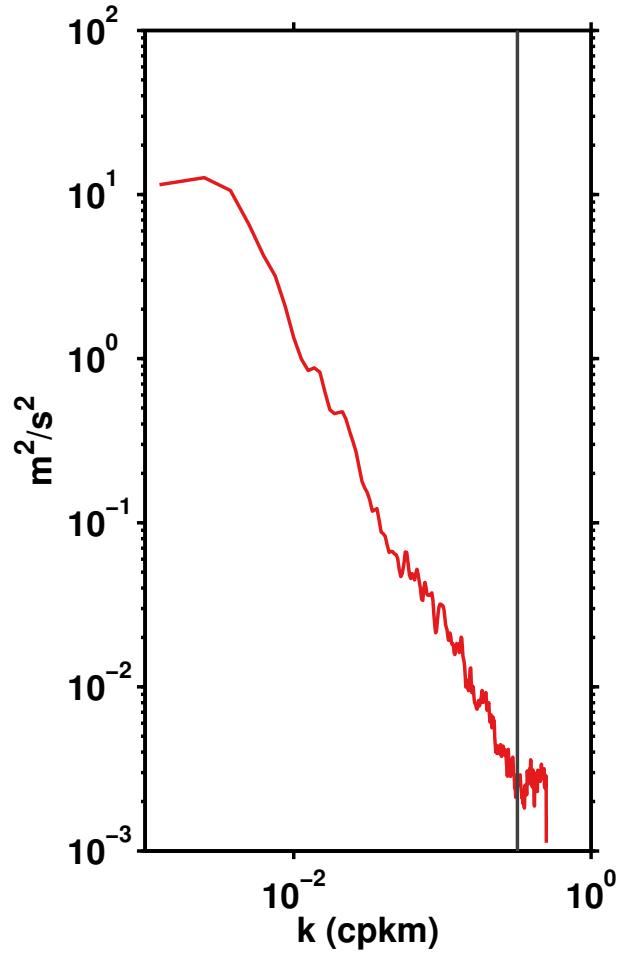

Fig. A2. $E_T = 0.5 \times (E_L + E_C)$ (red; total spectra) averaged over 24 bins (12–54 *m* depth) for SN100. Grey line denotes the horizontal wavenumber where the noise stops.

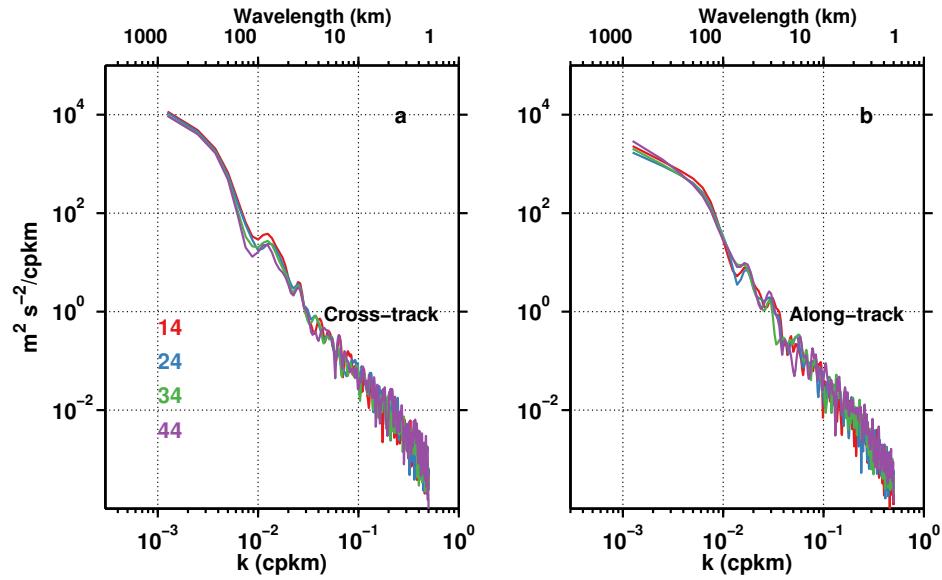

Fig. A3. (a) Cross-track and (b) along-track spectra at different depths for SN115. The depths at which the spectra are presented are 14, 24, 34 and 44 *m*.

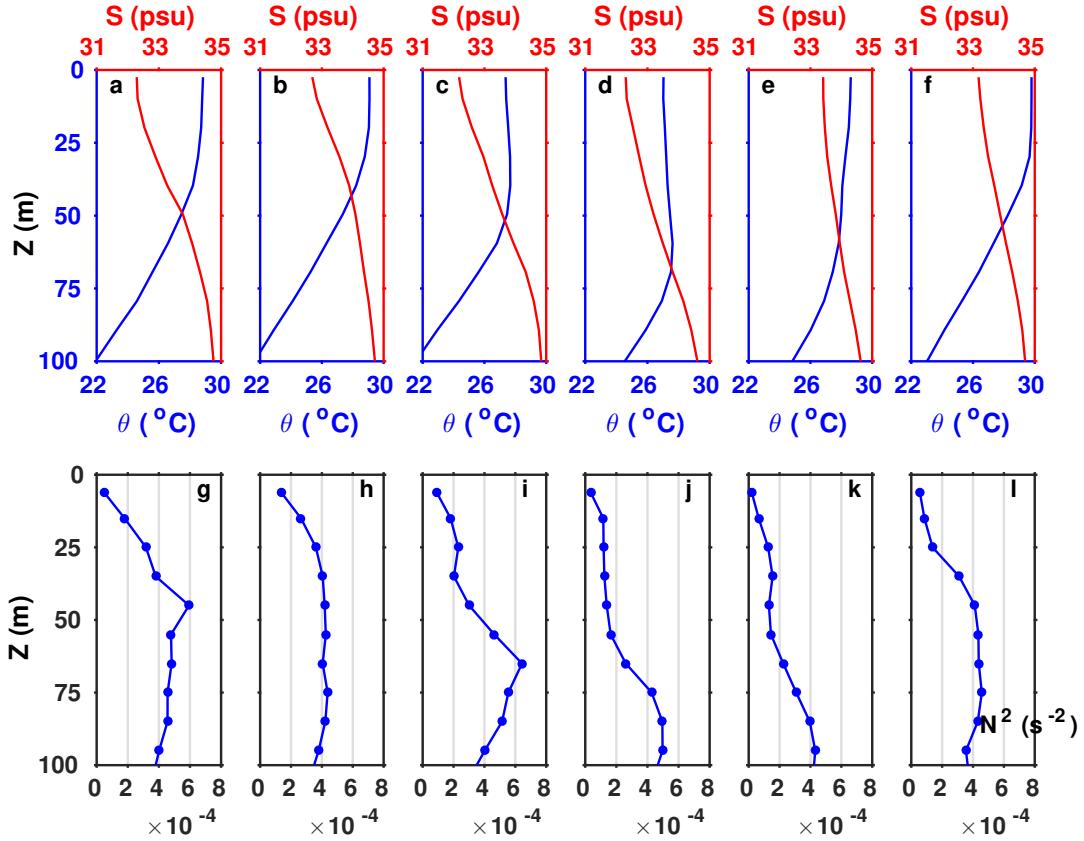

Fig. A4. ARGO profiles of temperature and salinity (first row) and the buoyancy frequency (second row) near the cruise tracks. The mean buoyancy frequency, when averaged over the depth of the current data (12-54 $m$), from these measurements is approximately $2 \times 10^{-4} s^{-2}$. Panels (a, g), (b, h), (c, i), (d, j), (e, k), and (f, l) represent cruise number SN88, SN100, SN110, SN112, SN115, and SN120 respectively.

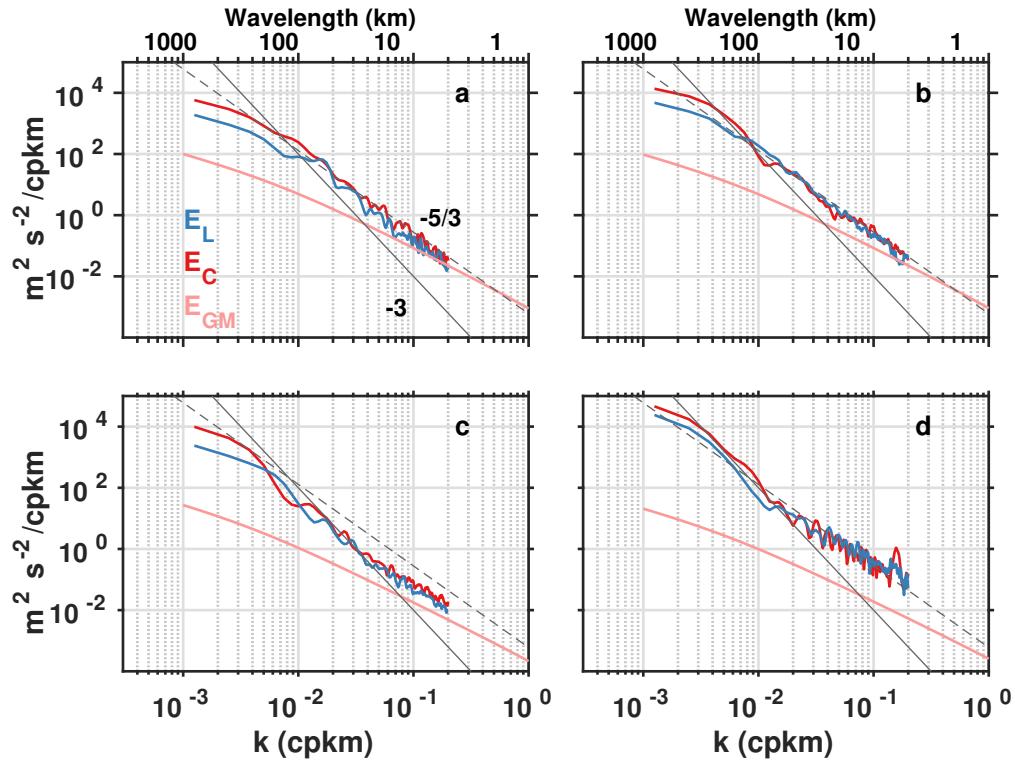

Fig. A5. Along-track ($E_L$; blue), cross-track ($E_C$; red) HKE spectra, and an estimate of the Garret-Munk ($E_{GM}$; pink) spectrum for (a) SN88, (b) SN100, (c) SN115, and (d) SN120. The slanted gray line represents the slope of $-3$ and the slanted gray dashed line represents the slope of $-5/3$.

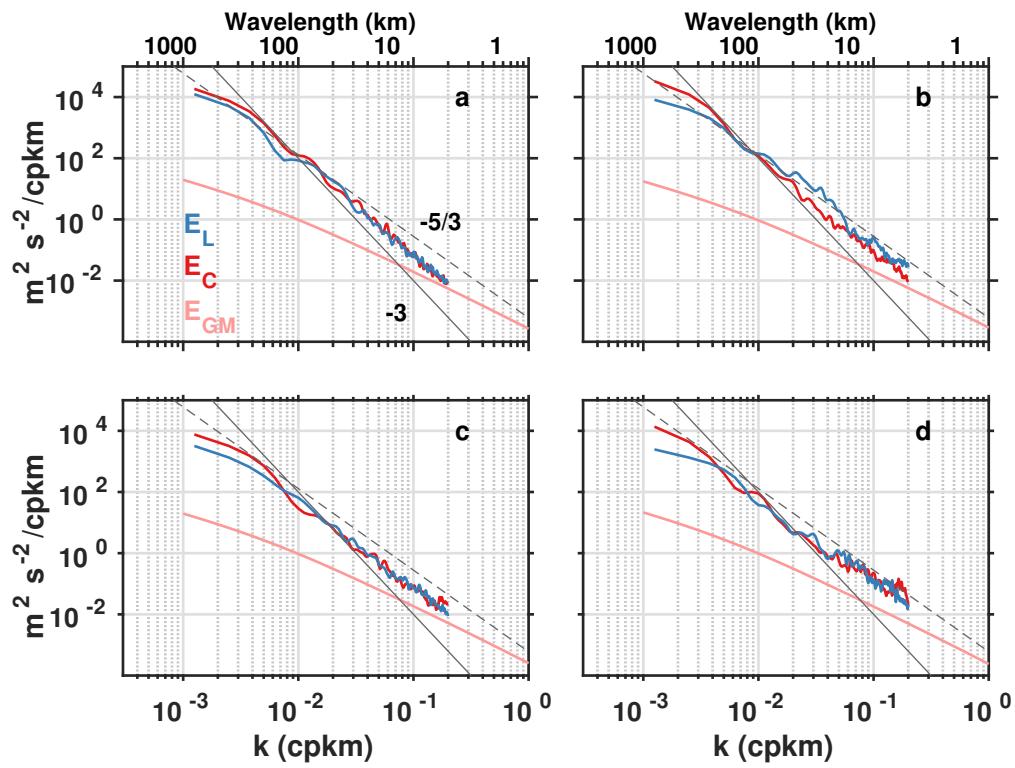

Fig. A6. Same as Figure A5 but for (a) SN110A, (b) SN110B, (c) SN112A, and (d) SN112B.